\documentclass[10pt]{article}

\usepackage[centertags]{amsmath}

\newcommand{\sect}[1]{\setcounter{equation}{0}\section{#1}}

\def\spazio#1{\vrule height#1em width0em depth#1em}

\def\ta{\Bigl(\Bigr.}
\def\tc{\Bigl.\Bigr)}

\begin{document}

\title{\bf  Two fermion relativistic bound states: hyperfine shifts.}

\author{R.Giachetti$\,{}^{1,2}$,
E.Sorace$\,{}^{2}$}

\date{}

\maketitle

\centerline{{\small ${ }^1$ Dipartimento di Fisica, Universit\`a di
Firenze, Italy. }}
\centerline{{\small  ${ }^2$ INFN Sezione di Firenze}}

\begin{abstract}
{ We discuss the hyperfine shifts of the Positronium levels in a
relativistic framework, starting from a two fermion
wave equation where, in addition to the Coulomb potential, the
magnetic interaction between spins is described by a Breit term.
We write the system of four first order differential equations describing
this model. We discuss its
mathematical features, mainly in relation to possible
singularities that may appear at finite values of the radial
coordinate. We solve the boundary value problems both in the
singular and non singular cases and we develop a perturbation
scheme, well suited for numerical computations, that allows to
calculate the hyperfine shifts for any level, according to well established
physical arguments
that the Breit term must be treated at the first perturbative order. We discuss our
results, comparing them with the
corresponding values  obtained from semi-classical expansions. }
pacs{03.65.Pm, 03.65.Ge}
\end{abstract}

\bigskip
\bigskip


\maketitle

\bigskip

%
%
%
%


\sect{Introduction} \label{Sec_introduction}


The relativistic description  of the fine structure of the hydrogen atom levels was
first proposed by Darwin \cite{Darwin} in a semi-classical treatment of the Dirac equation. This was
immediately followed by the Breit proposal of a two body
relativistic equation \cite{Breit} that, in addition to the Coulomb potential,  included a
quasi-static magnetic term  where the velocities were
substituted by the Dirac $\overrightarrow{\alpha}$-matrices according to a
proposal of Heisenberg. Shortly later Fermi calculated the
spectrum of a Dirac electron interacting with a Pauli nucleus and
deduced the values of nuclear magnetic momenta from the measured
hyperfine splitting \cite{Fermi}, using the Schr\"odinger non relativistic wave
functions to calculate the averages near the origin. In fact, following a Pauli suggestion, 
also the Breit paper contained a
semi-classical expansion along the lines of the Darwin work,
but eventually Breit himself was doubtful about the
actual predictivity of his equation and two years later he
reached the conclusion that the Coulomb
interaction should be treated exactly, while the magnetic term had
to be considered as a first order self-consistent perturbation, so
that only its diagonal matrix elements on the exact Coulomb
eigenstates were brought to bear. Since then, major achievements
were obtained by Pirenne \cite{Pirenne}, Berestetski and Landau
\cite{Landauetal}. Still using  the Breit semi-classical
approximation and the non relativistic Coulomb wave functions,
they were able to produce analytical approximations for the shifts
of the Parapositronium levels and for the ground state of the
Orthopositronium. In their approach they also included  the
annihilation term. In the first fifties, the appearance of the
Schwinger \cite{Schwi} and of the Bethe-Salpeter equations
\cite{BetheSalp,Salp} made it possible to calculate up to the
order $\alpha^5$ ($\alpha$ = fine structure constant) the shifts
of the first excited levels of the Positronium, by accounting for the effects of one and two virtual photons, self
energy and vacuum polarization, but still keeping the semi-classical perturbative
scheme \cite{KK,FM}. Later, in the seventies, due to the great improvement in
the experimental analysis of the atomic spectra, \cite{Berko}, to
the qualitative changes in the mathematical and physical framework
of symmetries and mainly to the new ideas of hadrons as composed by
quarks, the interest for the bound states and for the relativistic wave equations raised up
again and never weakened since,  \cite{Lepage}-\cite{Karsh}. The
challenge of the completely relativistic calculation of the hyperfine splitting 
was again pushed on foreground and various models for 
two spin $\frac 12$ interacting particles were proposed, with special
attention to the Positronium that constituted an ideal system both from a
theoretical and an experimental point of view: it appears, however,
that the use of the semi-classical expansion and non relativistic Coulomb wave
functions as a starting point has maintained his role. Even in a
later paper \cite{Krollinowski}, where  a non
perturbative treatment is claimed, the solutions of the Breit
equation are calculated analytically but expanding the equation up
to  the second order in $\alpha$. This latter procedure, in particular, may be
assumed somewhat safely in atomic physics, where the velocities are
of the same order of the fine structure constant, 
so that the expansion in $\alpha$ in fact corresponds to a
semi-classical approximation.
It becomes less and less justified in view of an
extension of the method to quark bound states (see \textit{e.g.}
\cite{Peshkin,Hardekopf}), where 
an expansion in the coupling constant is not allowed
and where it would be worth dealing from the
very beginning with relativistic states.

In the present paper we fill this old gap and we
present a completely relativistic treatment of the hyperfine
splitting based on the Breit approach, providing an effective
method for its computation for any spectral level. We thus analyze 
an approximate interaction within the framework of an exact kinematics.
It is rather evident that analytic results will be possible only for
the initial steps of our treatment, \textit{i.e.} for establishing the 
system of equations to be discussed and their reduction due to conserved
quantities. Analytic expressions will also be available
when looking for series or asymptotic solutions, but finding the spectrum
will necessarily be achieved by numerical methods.
Moreover we shall omit from our treatment the annihilation term.

The starting point is thus the two fermion relativistic wave
equation we presented in \cite{RGES}. In that paper, using a
canonical reduction of the relativistic kinematics of the two body
problem, we introduced Lorentz invariant interactions dependent
upon the reduced coordinates and we gave a solution of the
relative time problem. We then quantized the model assuming a
fermion nature for both of the two particles and we deduced a
completely Lorentz covariant internal dynamics, that was reduced
by separating the radial part in a multipole scheme exploiting
the conservations of angular momentum and parity. A further
reduction of the radial equations produced a final linear
system of order four containing the spectral
parameter to be determined. We proved that the Dirac and the
non-relativistic limits were recovered and we compared our model
with other existing models, \cite{Chi,SSM,Dare,Sazd}, finding a
general agreement.  The complete spectral curves from Dirac to
Positronium for pure Coulomb interaction were plotted for ground
and higher states and the crossing of terms inferred in \cite{SSM} was
precised and made concrete. In that paper we also looked at the possibility of
using the eigenfunctions of the degenerate singlet-triplet ground
states of the Positronium to set up a direct perturbative
calculation of the hyperfine splitting, using the usual
perturbative terms for the magnetic interaction: the answer was
negative even for the ground state, since the behavior of the
relativistic wave functions in the origin was not compatible with the magnetic interaction
terms obtained by the semiclassical approximation.

In order to investigate the hyperfine structure, we shall
therefore add to our pure Coulomb wave equation a Breit term in a
Lorentz covariant way. The radial reduction of the problem is again done
by using the still holding conservations of angular momentum and parity 
and here also the final wave equation 
reduces  to a fourth order linear system. Unfortunately the
situation is now less nice than in the pure Coulomb case, since a
singularity of the wave equation appears at finite positive values of the radial
coordinate. This occurrence had already been notified (see
\cite{OrazioWa}) and, to our knowledge, has prevented
up to date any correct integration attempt. However a careful
investigation shows that the behavior of the system is not so bad.
Indeed the singularity can be ``bridged'' and the spectrum can be
determined although, as previously said, only the first
perturbative order in the Breit interaction makes sense and higher
order corrections should be dealt with using the QED, \cite{Karsh}.
Therefore we are going to present a perturbative approach that
allows the complete numerical calculation of the contribution to the shift of any
level of the Para and the Orthopositronium due to the Breit term.
We shall show that in almost any case the presence of the additional singularity can
be avoided: only for the levels with odd parity and angular momentum $j=0$
this is impossible and the analysis has to be refined. However,
in order to provide numerical evidence of the correctness
of the Breit argument concerning the perturbative nature of the
magnetic interaction, we further  investigate the singular cases connected to the ground
states. A numerical approach based on a different
starting point and on Pad\'e approximation techniques has been
successful in calculating the shifts of the Parapositronium singlets
to the order $\alpha^4$, \cite{Orazio}. 
The principal merit of our scheme from a numerical
point of view is that we are able to obtain the results
uniquely by means of the numerical values of spectral levels, without any
need of using the eigenfunctions in order to calculate the matrix elements for the Breit term:
in the full relativistic case this 
leads to a faster and more precise approach for finding the spectrum, especially for the
Orthopositronium triplets. When comparing our results 
to the values of the hyperfine levels computed in the
semi-classical scheme and expressed in terms of simple fractions of powers of
$\alpha^2$, \cite{FM,AhB}, we find an excellent agreement up to the order $\alpha^4$. 
This is rather remarkable, 
since the pure Coulomb contribution -- where the correct kinematics and the 
recoil effects are exactly accounted for -- and the magnetic perturbative terms 
are here separately different with respect to 
the corresponding semi-classical contributions (more details are given in Section IV).
We want to stress that, apart some technical obstacles, our
general method has a great conceptual simplicity:
besides  its applications to any excited level of electromagnetic bound states with components of
arbitrary mass ratio, we believe that it should also be relevant
for models of different nature, for which non-relativistic
calculations are less reliable.

The paper is organized as follows. In Section II we
briefly recall the derivation of the wave equation for the two
fermion system with a general radial interaction and we present
the new equation with the additional  Breit term together with its
reduction that defines the spectral problem to be solved. The
procedure is completely parallel to that of the pure Coulomb
interaction explained in the previous paper \cite{RGES}, which we
refer to for detailed expressions, as, for instance, the explicit
form of the even and odd state vectors. 
The nature of the mathematical problem and the numerical methods
we have used for its solution are explained in  Section III. 
Here we also investigate the possible presence of singularities at a
non vanishing finite value of the radial coordinate and we comment on their properties.
Finally, in Section IV, we present
the results and we discuss them. A revisitation of the
perturbative expansion as we have used it in our computations (up to second order in the proof, but
evidently valid at any order) is given in the Appendix.

In the following we use units  such that $\hbar=c=1$.


\sect{The system for the Coulomb and Breit interaction}
\label{CoulombBreit}

In the paper \cite{RGES} we have described the method for
obtaining a Lorentz covariant wave equations for two fermions
interacting by means of a scalar potential. Obviously we
cannot reproduce the derivation here and we have to refer to
\cite{RGES} for details. It is however necessary to recall the
main definitions in order to have a minimum of self consistency of
the treatment.

\medskip
\noindent
\textbf{(\textit{a}) \textit{The  canonical variables.}}
\medskip

Denote by  $x^\mu_{(i)}$ and $p^\mu_{(i)}$ the Minkowski
coordinates and the momenta of two pointlike fermions with masses
$m_ {i}$, $i=1\,,\,2$. Define the tensor 
\begin{eqnarray}
\label{varepsilon_mu_nu}
\varepsilon_a^\mu(P)=\eta_a^\mu-\displaystyle{\frac{P_a\,[\,P^\mu
+\eta_0^\mu\sqrt{P^2}\,]}{\sqrt{P^2}\,[\,P_0+\sqrt{P^2}\,]}}\,,~\,
\varepsilon_0^\mu(P)=\,P^\mu/\sqrt{P^2}\cr
\end{eqnarray}
where $P^\mu=p^\mu_{(1)}+p^\mu_{(2)}$ is the total
momentum. It satisfies the identities
\begin{eqnarray}
\!\!\!\!\! \eta_{\mu\nu}\,\varepsilon_\alpha^\mu(P)\,
\varepsilon_\beta^\nu(P)\,= \,\eta_{\alpha \beta}\,,~~~
\eta_{\alpha \beta}\,\varepsilon_\alpha^\mu(P)\,
\varepsilon_\beta^\nu(P)\,= \,\eta^{\mu\nu}
\end{eqnarray}
and  therefore it represents a Lorentz transformation to the
$\overrightarrow{P}=0$ reference frame. We can use (\ref{varepsilon_mu_nu}) for the
construction of a canonical transformation to the variables
\begin{eqnarray}
&{}& \!\!\!\!\!\! Z^\mu=X^\mu+ \displaystyle{\frac
{\varepsilon_{abc}\,P_a\, \eta_b^\mu\, L_c}{\sqrt{P^2}\,[P_0+
\sqrt{P^2}]}}
+\displaystyle{\frac{\varepsilon_a^\mu}{\sqrt{P^2}}}\ta
q_a\,\breve r-{r_a\,\breve q}\tc
+\displaystyle{\frac{P^\mu}{P^2}}\,\breve q\,\breve r
\spazio{1.4}\cr &{}& \!\!\!\!\!\! \breve{q}=\varepsilon^\mu_0\,
q_\mu\,,~~~\, \breve{r}=\varepsilon^\mu_0 r_\mu\,,~~~
q_a=\varepsilon_a^\mu\, q_\mu\,,~~~ r_a=\varepsilon_a^\mu r_\mu
\label{variabili}
\end{eqnarray}
where 
\begin{eqnarray}
\label{Xpr}
X^\mu=\frac 12 \,\Bigl(x^\mu_{(1)}+x^\mu_{(2)}\Bigr)\,,~~~~~~~
r^\mu\,=x^\mu_{(1)}-x^\mu_{(2)}\,,~~~~~~~q^\mu=\frac 12 \,\Bigl(p^\mu_{(1)}-p^\mu_{(2)}\Bigr)\,.
\end{eqnarray}
Both $r_a$ and $q_a$ are Wigner vectors of spin one, as well as ${Z_a}$
is a Newton-Wigner position vector for a particle with angular
momentum ${L_a}=\varepsilon_{abc}\,\,r_b\, q_c\,$. In terms of
(\ref{variabili}), the two particles mass shell conditions  $
p_{(i)}^2=m_{i}^2$ can be put into the form 
\begin{eqnarray}
\ta\, q_aq_a+m_{1}^2\,\tc ^{1/2}+\ta \,q_aq_a+m_{2}^2\,\tc
^{1/2}\, =\lambda\,,~~~~~~~~~~~
 \lambda \,\,\breve{q}=\frac 12 (m_{1}^2-m_{2}^2)\,,
\label{mass_shell}
\end{eqnarray}
where $\lambda=\sqrt{P^2}$ while the variable $\breve{q}$
can be fixed and generates a canonical reduction of the phase
space. Its conjugate coordinate, namely the relative time
coordinate  $\breve{r}$,  is cyclic and becomes a kind of
a gauge function that is chosen {\it a posteriori} in order to
recover the complete Minkowski description for each of the two
particles. For instance, a useful choice could be $\breve{r}=0$,
although there is no necessity of requiring  such condition.
 It is now straightforward introducing a Lorentz covariant
interaction by changing the mass shell condition
(\ref{mass_shell}) with the addition of a scalar potential
depending upon the Lorentz invariant relative separation
$~r=(r_ar_a)^{1/2}$. Thus a relativistic two-body system
interacting by means of a potential $V(r)$ is described by
(\ref{mass_shell}) where $\lambda$ will be substituted by
$h(r)=\lambda-V(r)$.

\medskip
\noindent \textbf{(\textit{b}) \textit{The  two fermion quantum
system.}}
\medskip

Let us now quantize the equation (\ref{mass_shell}) assuming that
both particles are fermions.
In order to determine the structure of the wave equation, it is sufficient to
consider the case of two free fermions, since the interaction will
then be introduced in the way previoulsy described. The two particle wave function is simply the
tensor product of the two single particle wave functions, and must
therefore  satisfy the two separate Dirac equations determined by the operators
$D_1=(\frac 12 P_\mu+q_\mu)\, \gamma_{(1)}^\mu-m_1$ and
$D_2=(\frac 12 P_\mu-q_\mu)\,\gamma_{(2)}^\mu-m_2$, where we 
adopt the notation $\gamma_{(1)}^\mu=\gamma^\mu\otimes{\bf
I}_4$, $\gamma_{(2)}^\mu={\bf I}_4\otimes\gamma^\mu$  and where
${\bf I}_4$ is the unity matrix in four dimensions. Introducing
the new set of $\gamma$-matrices
$\breve{\gamma}\equiv\breve{\gamma}(P)=\varepsilon_0^\mu(P)\gamma_\mu$,
$\gamma_{a}\equiv\gamma_{a}(P)=\varepsilon_a^\mu\gamma_\mu$  and
using the canonical operators corresponding to the variables
(\ref{variabili}), we easily find 
\begin{eqnarray}
\lambda=q_a\Bigl( \breve{\gamma}_{(1)}\,{\gamma_{(1)}}_{a}-
\breve{\gamma}_{(2)}\,{\gamma_{(2)}}_{a}\Bigr)+\breve{\gamma}_{(1)}
m_1+ \breve{\gamma}_{(2)}m_2\,,~~~~~~~~~~
 \lambda
\,\,\breve{q}=\frac 12 (m_{1}^2-m_{2}^2)\,. \label{Diracnew}
\end{eqnarray}
The system (\ref{Diracnew}) has
exactly the same content as the initial system of the two
independent Dirac equations. However, put in this form, we see
that the variable $\breve{q}$ remains fixed, in complete agreement
with the classical canonical reduction. The cyclic character of
$\breve{r}$ is also evident from the Lorentz scalar identity for
the phase of plane waves ${p_{(1)}}^\mu\,
{x_{(1)\,\mu}}+{p_{(2)}}^\mu \,{x_{(2)\,\mu}} =P^\mu
Z_\mu-q_a\,r_a\,$. Moreover the definition of
$\breve{\gamma},\gamma_{a}$ is  actually a unitary transformation
on the $\,\gamma^\mu\,$ 4-vector. Hence, as long as $P$ is
conserved, the matrices  can be represented by the usual $\gamma$
matrices.
The first of equations (\ref{Diracnew}) is  thus the
Lorentz-invariant equation for the two-fermion free system. Its
sixteen eigenvalues are immediately calculated, yielding the
expected four singular values 
\begin{eqnarray}
\lambda=\,\pm({q_a\,q_a+m_1^2})^{1/2}
\pm({q_a\,q_a+m_2^2})^{1/2}\,,~
\pm({q_a\,q_a+m_1^2})^{1/2}
\mp({q_a\,q_a+m_2^2})^{1/2} \,, \label{singval}
\end{eqnarray}
each singular value having multiplicity four.

The equation for the interacting system is simply obtained by substituting $\lambda$ with
$h(r)=\lambda-V(r)$ in the first of the (\ref{Diracnew}) equations. 
For future discussions,  we find it
useful to introduce the mass parameters $M=m_1+m_2\,,\,\,
\mu=m_1-m_2\,,\,\,\rho=\mu/M\,\,$ and to reorder the basis of the
states making a linear transformation on the tensor product of the
two spinor spaces  such that the system at rest is diagonalized
and the four singular values (\ref{singval})  are put in the order
$M$, $-M$, $-\mu$ and $\mu$. We then make a further linear
transformation such that, in each four dimensional eigenspace
corresponding to the above energy eigenvalues, the square and  the
third  component of the total spin  $S={\mathbf I}_4\otimes {\mathbf
\sigma} + {\mathbf \sigma}\otimes {\mathbf I}_4$ (where ${\mathbf \sigma}$ is
the Dirac spin) are also diagonalized and ordered with the triplet
always following the singlet.

We finally remind that the global parity transformation is given
by the product of orbital and internal parity transformations. In
our picture the internal parity is 
$\breve\gamma\otimes\breve\gamma={\mathrm{diag}}({\bf I}_8,-{\bf
I}_8)$.
It can be verified that the global angular momentum
${\mathbf J}={\mathbf L}+{\mathbf S}$ and the parity are conserved, so that, together
with $\lambda$, they provide a classification of the states of the
global symmetry.

\medskip
\noindent
\textbf{(\textit{c}) \textit{The  radial equations}}
\medskip

In the basis we
have chosen, the free Hamiltonian operator $H_0$ is a $16\times
16$ matrix of the form 
\begin{eqnarray}
H_0 =\left(\,
\begin{matrix}
 {\cal J}_M & {\cal H}_0 \spazio{0.8}\cr
 {\cal H}_0  & {\cal J}_\mu\cr
\end{matrix}
\,\right) \,, \label{H}
\end{eqnarray}
where ${\cal J}_\Lambda=\Lambda\,{\mathrm{diag}}({\bf I}_4,-{\bf
I}_4)$, $\Lambda=M,\mu$ and ${\cal H}_0$ is a $8\times 8$ matrix
whose elements are spherical differential operators (see
\cite{RGES} for the explicit form). We next construct the ``even''
and ``odd'' states $\Psi_{\pm}$  with  assigned angular momentum
$(j,m)$ and given parity $(-)^j$ or $(-)^{j+1}$, whose 16
components are collected in 4 groups indexed by the eigenvalues
$\pm\Lambda$  of the free system at rest. 
\begin{eqnarray}
\Psi_{\pm}={}^t\!\Bigl(\,
\Psi_{\pm}^{(M)},\, \Psi_{\pm}^{(-M)},\, \Psi_{\pm}^{(-\mu)},\,
\Psi_{\pm}^{(\mu)}
\,\Bigr)\,.
\label{psiMmu}
\end{eqnarray}
In each
group the components are singlet-triplet ordered, namely 
\begin{eqnarray}\Psi_{\pm}^{(\Lambda)}={}^t\!\Bigl(\,
\psi_{\pm,0}^{(\Lambda)}\,,\, \psi_{\pm,1_+}^{(\Lambda)}\,,\,
\psi_{\pm,1_0}^{(\Lambda)}\,,\, \psi_{\pm,1_-}^{(\Lambda)}
\,\Bigr)\,,
\label{psipm}
\end{eqnarray}
where the subscript ``$0$'' refers to the singlet component, while
``$1_+,1_0,1_-$''denote the triplet components. The explicit
expressions have been determined using the standard algorithms of
the composition of angular momenta by means of Clebsch-Gordon
coefficients and are therefore linear combinations of spherical
harmonics. They are fully reported in \cite{RGES}.

By applying the Hamiltonian operator (\ref{H}) to the states
(\ref{psiMmu}) we obtain a system of radial differential equations
by requiring the vanishing of the coefficients of the different
spherical harmonics in each component of the resulting vector.
Indeed we get a large number of differential equations
(\textit{e.g.} 34 when starting with $\Psi=\Psi_{+}$ ), but of
course, as one should expect, only eight of them are independent
for each state with definite parity. Moreover a closer look to
them shows that four of these eight equations are algebraic
relations that can be used, by an appropriate choice of the
unknown functions, to obtain a system of only four differential
equation for each of the two state vectors $\Psi_{+}$ and $\Psi_{-}$. Finally,
adding a Lorentz invariant interaction as specified in item
$(a)$, the fourth order system reads
\begin{eqnarray}
{\displaystyle \frac {dY(r)}{dr}}+{\cal M}\,Y(r)=0\,,
\label{sistema_mat}
\end{eqnarray}
where $Y(r)={}^t\!(y_1(r),y_2(r),y_3(r),y_4(r))$ and ${\cal
M}$ is a matrix with general structure
\begin{eqnarray}
{\cal M}= \left[ {\begin{array}{cccc} \phantom{-}0~ &
\phantom{-}E(r)~ & \phantom{-}F(r)~ & \phantom{-}0~
\spazio{1.2}\\
\phantom{-}E(r)~ & \phantom{-}{\displaystyle \frac {1}{r}}~  &
\phantom{-}0~ &  - F(r)~ \spazio{1.2}\\ [2ex] \phantom{-}{G}(r)~ &
\phantom{-}0~ & \phantom{-} {\displaystyle \frac {2}{r}}~  &
\phantom{-}E(r)~ \spazio{1.2}\\ [2ex] \phantom{-}0~ &  - {G}(r)~ &
\phantom{-}E(r)~ & \phantom{-}{\displaystyle \frac {1}{r}}~
\end{array}}
 \right]
\label{sysmat}
\end{eqnarray}
The explicit expressions of $E(r)$, $F(r)$ and $G(r)$ for
the even case are the following ones
\begin{eqnarray}
E(r)={\displaystyle \frac {\sqrt{j(j + 1)}\,\mu
}{r\,{h}(r)}}\,, ~~~
F(r)= {\displaystyle \frac {\mu^2-h^2(r)}{2h(r)}} \,,~~~
G(r)={\displaystyle \frac {h(r)}{2}}\,\Bigl(1 -
{\displaystyle \frac {r^{2}\,M^{2} + 4\,j\,(j + 1)}{r^{2}\,
{h}^2(r) }}\,\Bigr)\, \label{equaUVWeven}
\end{eqnarray}
and they specialize to the Coulomb interaction when
${h}(r)=\lambda+\alpha /r\,$, $\alpha$ being the fine structure
constant.

As explained in \cite{RGES}, the odd coefficients are obtained
from the previous ones by a parity transformation, whose action
simply results in the change $M\rightarrow -\mu$ and
$\mu\rightarrow -M$.  In \cite{RGES} it has also been shown
how the Dirac and the non relativistic limits are recovered from
(\ref{sysmat}). We finally observe that
for $E(r)=0$ (\textit{e.g.} for $j=0$ or for $\mu=0$ in the even
case) the fourth order system splits into separate second order
subsystems, making it easier the numerical solution of the
spectral problem. In the general case, however, the complete
fourth order system has to be considered.

\medskip
\noindent
\textbf{(\textit{d}) \textit{The  addition of the Breit term}}
\medskip

The spin-spin interaction can be described by introducing  a Breit
term, in addition to the Coulomb interaction, in the relation
(\ref{Diracnew}) that becomes
\begin{eqnarray}
&{}&\!\!\!\!\!\!\! \lambda+\frac\alpha r\Bigl[1- 
 \frac
12\Bigl(\breve{\gamma}_{(1)}{\gamma_{(1)}}_{a}\,\breve{\gamma}_{(2)}{\gamma_{(2)}}_{a}
+(\breve{\gamma}_{(1)}{\gamma_{(1)}}_{a}\,\frac{r_a}r)\,(\breve{\gamma}_{(2)}{\gamma_{(2)}}_{b}\,
\frac{r_b}r)
\Bigr)\Bigr] \spazio{1.0}\cr 
&{}&\qquad\qquad\qquad\qquad =\,q_a\Bigl(
\breve{\gamma}_{(1)}{\gamma_{(1)}}_{a}-
\breve{\gamma}_{(2)}{\gamma_{(2)}}_{a}\Bigr)+\breve{\gamma}_{(1)}
m_1+ \breve{\gamma}_{(2)}m_2. \label{BreitHam}
\end{eqnarray}
The Hamiltonian matrix must be modified with
respect to (\ref{H}) in order to account for the presence of the
additional $\gamma$-matrices. This modification
as well as the change of the basis are straightforward. The
angular momentum and parity are conserved and
the radial equations are again deduced by applying the new
hamiltonian to the states $\Psi_{\pm}$, as in the pure Coulomb
case. The general features of the systems of radial equations are
also preserved and the final result is again a fourth order linear system.
In the even case and with $h(r)=\lambda+\alpha/r$,  the matrix ${\cal B}$ of this system
reads 
\begin{eqnarray}
{\cal B}= \left[ {\begin{array}{cccc} \phantom{-}0~ &
\phantom{-}E(r)~ & \phantom{-}F(r)~ & \phantom{-}0~
\spazio{1.2}\\
\phantom{-}{E}_\varepsilon(r)~ & \phantom{-}{\displaystyle \frac
{1}{r}}~  & \phantom{-}0~ &  {F}_\varepsilon(r)~ \spazio{1.2}\\
[2ex] \phantom{-}{G_{1,\varepsilon}}(r)~ & \phantom{-}0~ &
\phantom{-} {\displaystyle \frac {2}{r}}~  &
\phantom{-}{E}_\varepsilon(r)~ \spazio{1.2}\\ [2ex] \phantom{-}0~
&  {G_{2,\varepsilon}}(r)~ & \phantom{-}E(r)~ &
\phantom{-}{\displaystyle \frac {1}{r}}~
\end{array}}
 \right]
\label{MatBreit}
\end{eqnarray}
where $E(r)$ and $F(r)$ are given in (\ref{equaUVWeven}).
The remaining matrix elements read 
\begin{eqnarray}
&{}& \!\!\!\!\! E_\varepsilon(r)={\displaystyle \frac {\sqrt{j(j +
1)}\,\mu }{r\,{h}(r)-2\alpha\varepsilon}}, ~~~~~~~ F_\varepsilon(r)=
{\displaystyle \frac
{(h^2(r)-\mu^2)r^2-(2\alpha\varepsilon)^2}{2r(rh(r)-2\alpha\varepsilon)}}
\spazio{1.0}\cr &{}& \!\!\!\!\!
G_{1,\varepsilon}(r)={\displaystyle \frac {h(r) }{2}}    +
{\displaystyle \frac {4\,\alpha \,\varepsilon  + {\displaystyle
\frac {4\,j\,(j + 1)}{2\,\alpha \,\varepsilon  - r \,h(r)}}  +
{\displaystyle \frac {r^{2}\,M^{2}}{4\,\alpha \,\varepsilon  - r\,
h(r)}} }{2\,r}} \spazio{1.0}\cr &{}& \!\!\!\!\!
G_{2,\varepsilon}(r)={\displaystyle \frac {2\,j\,(j + 1)}{r^{2}\,
h(r)}}  + {\displaystyle \frac {4\,\alpha ^{2}\,\varepsilon ^{2} +
( -
 h(r)^{2} + M^{2})\,r^{2}}{2\,r\,
 (r\,h(r) - 2
\,\alpha \,\varepsilon)}} \label{equaUVWeven2}
\end{eqnarray}

Again the matrix of the odd system is obtained by changing
$M\rightarrow -\mu$ and  $\mu\rightarrow -M$.  It is also
immediate to verify that for $\varepsilon=0$ (\ref{MatBreit})
reduces to (\ref{sysmat}).

Some remarks are in order. In the first place, we observe that in
the matrix (\ref{MatBreit}) we have introduced a parameter
$\varepsilon$, not present in (\ref{BreitHam}) and reproducing
the latter for the $\varepsilon=1/2$. We shall see in
the following that an appropriate use of this parameter permits the
calculation of the first perturbative terms in the Breit
interaction in a way that is numerically much more efficient than
the usual computations by means of the eigenfunctions. Secondly in
(\ref{BreitHam}) and therefore in the system produced by
(\ref{MatBreit}) no anomalous magnetic moment is present. This
absence would constitute a drawback for the calculation of
the hyperfine structure of the hydrogen atom, but remains an
acceptable approximation for the positronium, which is the only
case we are going to consider later on. Finally, in the
even and odd systems describing the positronium, we will let
$\mu=0$ and $M=2m_{{\mathrm e}}$, where $m_{{\mathrm e}}$ is the
electron mass.


\sect{The numerical treatment of the Breit interaction}
\label{PerBreit}

We now discuss some general properties of the equations for the
positronium with a particular attention to the possible presence
of singularities other than the origin and the infinity. We find
it useful to introduce the new dimensionless independent variable
$x$ and the new eigenvalue $w$ as follows 
\begin{eqnarray}
x=m_{\mathrm{e}}r\,,~~~~~~~
w=\Bigl(\frac{m_{\mathrm{e}}}2\,\alpha^2\Bigr)^{-1}\,(\lambda - 2m_{\mathrm{e}})\,.
 \label{dimensionless}
\end{eqnarray}
The value of the fine structure constant has been assumed
as $\alpha=0.007297372568$, \cite{PartData}. In the following we
distinguish the discussion of the even case, that develops
essentially according to the classical O.D.E. theory \cite{CoddLev}, from the odd case, that
poses some new problems.

\medskip
\noindent
\textbf{(\textit{a}) \textit{The  even case}}
\medskip

We begin from the case with even parity. From (\ref{MatBreit}) we
see that ${\cal B}_{12}={\cal B}_{21}={\cal B}_{34}={\cal
B}_{43}=0$ when $\mu=0$. This means that the fourth order system
generated by (\ref{MatBreit}) decouples into two second order
systems for the unknown functions $(y_1(x),y_3(x))$ and
$(y_2(x),y_4(x))$ that, in turn, can be reduced to two second
order differential equations for $y_1(x)$ and $y_2(x)$. It has
also been shown in \cite{RGES} that for $j=0$ only the system for
$(y_1(x),y_3(x))$ makes sense, while for $j>0$ both systems
contribute to determining the levels. The first of these two
equations, introducing the unknown function $u(x)$ defined by
${y_{1}}(x)=[{(4 + \alpha ^{2}\,w)\,x+ 2\,\alpha
}]^{1/2}\,{x^{-3/2}} \,{u}(x)$, reads 
\begin{eqnarray}
&{}& \!\!\!\!\!\! {\frac {d^{2}}{dx^{2}}}\,{u}(x) + \Bigl[ -
{\displaystyle \frac {8\,\alpha \,\varepsilon }{(4+ \alpha
^{2}\,w)\,x - 2\alpha\,(4\, \varepsilon  - 1) }} - {\displaystyle \frac {3\,\alpha
^{2}}{(( 4+ \alpha ^{2}\,w)\,x + 2\,\alpha )^{2}\,x^2}}\spazio{1.2}\cr
&{}& -\frac 1{16 x^2} \Bigl( - \alpha ^{2}\,w\,(8+ \alpha ^{2}
\,w)\,x^{2} - 4\,\alpha \,(4+\alpha ^{2}\,w)\,(2\,\varepsilon  +
1) \,x - 4\alpha^2 (4\,\varepsilon  + 1)+
16\,j\,(j + 1)\Bigr) 
\spazio{1.2}\cr &{}& - {\displaystyle \frac {4\,\alpha
\,\varepsilon \,j\,(j
 + 1)}{((4+ \alpha ^{2}\,w)\,x - 2\alpha\,(2\,\varepsilon  - 1))
\,x^{2}}}\Bigr] {u}(x)=0 
\label{even1}
\end{eqnarray}
It can be seen that, in addition to the usual
singularities in the origin and at infinity, the equation presents a
new singularity in the point
$x=2\alpha\,(4\varepsilon-1)/(\alpha^2 w+4)$, that assumes finite
positive values for $\varepsilon>1/4$. Although we shall show that
solutions exist also for $\varepsilon>1/4$ -- in fact we shall give
an explicit solution for $\varepsilon=1/2$ --, in carrying out our
perturbative program we can avoid this singularity, as well as
almost all  those we shall encounter in later developments.
Indeed, according to the perturbative approach we are going to
explain here below -- and whose proof is given in Appendix A --,  it is
sufficient to solve the equation for values $\varepsilon<1/4$.
There is, however, one instance in the case with odd parity where the singularity must be
explicitly taken into account.

If  ${y_{2}}(x)=[{(4+\alpha ^{2}\,w)\,x + 2\, \alpha\, (
2\,\varepsilon +1)}]^{1/2}\,{x^{-3/ 2}}\,{z}(x) $, the second
equation of the even case in  $z(x)$ is 
\begin{eqnarray}
&{}&\!\!\!\!\!\!\!\!\!\!\! {\frac {d^{2}}{dx^{2}}}\,{z}(x) +
\Bigl[ - {\displaystyle \frac {8\,\alpha \,\varepsilon }{(
4+\alpha ^{2}\,w)\,x + 2\,\alpha \,(1 - 2\,\varepsilon )}}
- {\displaystyle \frac {3\,\alpha
^{2}\,(2\,\varepsilon
 + 1)^{2}}{((4+\alpha ^{2}\,w )\,x + 2\,\alpha \,(2\,\varepsilon
  + 1))^{2}\,x^{2}}}
\spazio{1.2}\cr &{}&  
- \frac 1{16 x^2}\Bigl(  - \alpha ^{2}\,w\,(8+\alpha
^{2}\,w)\,x^{2}
- 4\,\alpha \,(4+ \alpha ^{2}\,w)\,(2\,\varepsilon  + 1)\,x - 4\,(2\, \varepsilon  +
1)^{2}\,\alpha ^{2}  +16\,j\,(j +
1)\Bigr)
\spazio{1.2}\cr &{}& - {\displaystyle \frac {4\,\alpha\,\varepsilon \,j(j +
1)}{((4+\alpha ^{2}\,w)\,x + 2\,\alpha )\,x^{2}}} + {\displaystyle \frac {2\,\alpha
\,(2\,\varepsilon  + 1)}{((4+\alpha ^{2}\,w)\,x + 2\,\alpha \,(
2\,\varepsilon  + 1))\,x^{2}}}
  \Bigr]\,{z}(x)=0\,.
\label{even2}
\end{eqnarray}
This equation develops a new singularity at
$x=2\alpha(2\varepsilon-1)/(4+\alpha^2w)$, that assumes  positive
values for $\varepsilon>1/2$. This new singularity is therefore
not effective for the same reasons explained above. It would be
such also when considering the Breit term as non perturbative,
and it would only act through a modification of the singularity in the
origin.

The boundary value problem posed by (\ref{even1}) and
(\ref{even2}) for sufficiently small values of $\varepsilon$,
namely without the additional singularity, is very classical. Both
the boundary points, the origin and the infinity, are singular and
what we have to do for starting the numerical procedure, is to
determine the initial conditions in the neighborhood of those
points by looking for analytic approximations of the solutions in
the form of series or asymptotic expansions respectively. We then
apply the double shooting method in order to determine the value
of the spectral parameter with the desired accuracy. As a matter
of fact  both the series and the asymptotic expansions produce two
linearly independent solutions, but in each case only one solution
can be accepted: we are thus in the so called ``\textit{limit point}''
case of the Weyl classification of the boundary value problems
\cite{CoddLev} and therefore there is no ambiguity in choosing the
appropriate solution for giving the initial conditions and
starting the numerical integration, once  the spectral parameter
has been assigned a numerical value that will be adjusted in the
successive integrations until when the two solutions coming from
zero and infinity, as well as their derivatives, match within the
required accuracy.  This is actually the spectral condition that
simply reduces to checking  the equality of logarithmic
derivatives in a chosen crossing point, whose location is
immaterial for the result. The regular series solutions for the
first and second  equations of the even case are of the form
$y_{i}(x)=A_i \,x^{\rho_i}S_i(x)$, $i=1,2$,  where $A_i$ are
integration constants, $S_i(x)$ are power series in $x$ with
zeroth order term equal to unity and $\rho_i$ are the  positive
indices 
\begin{eqnarray}
&{}&
\rho_1={\displaystyle \frac {1}{2}}  + \frac 1{2\,(  1 -
2\,\varepsilon )}\,\Bigl[ (1 - 2\,\varepsilon )^{2}\,(4 - \alpha
^{2}\,(4\,\varepsilon  + 1))  + 4\,(1 -
2\,\varepsilon )\,j(j + 1)\Bigr]^{1/2} \spazio{0.8}\cr 
&{}&
 \rho_2={\displaystyle
\frac {1}{2}}  + \frac 12  \Bigl[   4\,(1+2\,\varepsilon )\,j(j +
1)- (1+2\,\varepsilon )^{2}\,\alpha ^{2}\Bigr]^{1/2}\,.
\end{eqnarray}
The asymptotic solutions for the two equations are of the
form $y_i(x)=B_i\, \exp[-\kappa_i x]\,x^{\nu_i}T_i(x)$, $i=1,2$,
where $B_i$ are again integration constants, $T_i(x)$ are power
series in $1/x$ with zeroth order term equal to unity,  while the
two positive numbers $\kappa_i$ and the two indices $\nu_i$
coincide for both equations and are equal to $\kappa$ and $\nu$,
where 
\begin{eqnarray}
&{}&\!\!\!\!\!\!\!\!\!\! {\kappa }=\frac 14 {\sqrt{ - w\,(\alpha ^{2}\,w
 + 8)}}\,,\spazio{1.2} \cr
&{}&\!\!\!\!\!\!\!\!\!\! {\nu}={\displaystyle \frac {\alpha ^{2}\,w\,(8+\alpha ^{2}\,w
 )\,(2\,\varepsilon  + 1) + 16}{2\,(4+\alpha ^{2}\,w )\,
\sqrt{ - w\,(8+\alpha ^{2}\,w )}}} \label{kappanupari}
\end{eqnarray}
We finally remark that only the first terms of the series
and asymptotic expansions have been determined analytically. The
following ones have been calculated by numerical codes, and the
number of terms to be taken has been chosen according to the
following criterion: in the point where the initial conditions for
the numerical integrations were assigned the result of the
substitution of the solution into the differential equation
divided by the solution itself had to be less that $10^{-15}$. We
also remark that tests have been made also for the arithmetic
precision of the calculations and the number of meaningful figures
has always been kept sufficiently high.

Let us now discuss the solution of equation (\ref{even1}) 
for the even ground state, $j=0$, with
$\varepsilon=1/2$, that presents a singularity at the positive
point $x_s=2\alpha/(4+\alpha^2 w)$. We can study the series
solutions in $x_s$ and we realize that this singular point is in
the case of the ``\textit{limit cycle}'' of the Weyl
classification \cite{CoddLev}, independently of the value of $w$
in the physical domain. This means that in the neighborhood
of $x_s$ the equation admits two finite solutions that can be used
for matching the solution and its derivative coming from the
origin to the solution and the corresponding derivative coming
from infinity, forming therefore what in mathematics is referred to as
a  ``classical solution'' of the differential equation. Procedures
of this type and also with more elaborate matching conditions have
been occasionally considered, see \textit{e.g.} \cite{Cheon}, but
usually more as an investigation of mathematical possibilities,
than under the real necessity of solving a specific problem.
The first terms of the solution in the neighborhood of $x_s$ are
\begin{eqnarray}
{y_{s}}(x)=A\,(x - {x_{s}})\,\Bigl(1 + {\displaystyle \frac
{2\,\alpha
 \,(x - {x_{s}})}{\alpha ^{2}\,w + 4}} \Bigr) + B\,\Bigl(1+{\displaystyle
\frac {4\,\alpha \,(x - {x_{s}})\,\mathrm{ln}( \left|  \! \,  x - {x_{s}}\, \!  \right|
)}{\alpha ^{2}\,w + 4}}  \Bigr)
\end{eqnarray}
The index in the origin is $\rho=\frac
12(1+\sqrt{4-3\alpha^2})$, while the parameters for the asymptotic solution are
those given in (\ref{kappanupari}) with $\varepsilon=1/2$. The results
will be discussed in the next section.

\medskip
\noindent
\textbf{(\textit{b}) \textit{The  odd case}}
\medskip

From the remarks following (\ref{MatBreit}), we see that the
matrix for the odd system is obtained by substituting $M=0$ and
$\mu=-2m_{\mathrm e}$ in  (\ref{MatBreit}). In the dimensionless
variables (\ref{dimensionless}), the explicit expression of the
system we get is 
\begin{eqnarray}
&{}& {\frac {d}{dx}}\,{y_{1}}(x) - {\displaystyle \frac {2
\,\sqrt{j(j + 1)}}{r\,{h}(x)}}\,{y_{2}}(x)  + {\displaystyle \frac
{1}{2}} \,{\displaystyle \frac {(4 - {h}^2(x))}{{h}(x)}} \,{y
_{3}}(x)=0
 \spazio{1.2}\cr 
&{}& 
{\frac {d}{dx}}\,{y_{2}}(x) -
{\displaystyle \frac {2 \,\sqrt{j(j + 1)}\,{y_{1}}(x)}{x\,{h}(x) -
2\,\alpha \,\varepsilon}}  +  \frac 1x y_{2}(x)  + 
{\displaystyle \frac
{1}{2}} \,{\displaystyle \frac {( - 4\, \alpha ^{2}\,\varepsilon
^{2} + ({h}^2(x) - 4)\,x^{2})\, {y_{4}}(x)}{x\,(x\,{h}(x) -
2\,\alpha \,\varepsilon )}} =0 
\spazio{1.2}\cr 
&{}& 
{\frac
{d}{dx}}\,{y_{3}}(x) + {\displaystyle \frac {1}{2}} \,
 \Bigl(   {h}(x) + \frac 1x\,(\, {\displaystyle {4\,\alpha \,
\varepsilon  + {\displaystyle \frac {4\,j\,(j + 1)}{2\,\alpha \,
\varepsilon  - x\,{h}(x)}} }} )   \Bigr) \,{y_{1}}(x) +  
 \frac 2x \,{y_{3}}(x)
- {\displaystyle \frac {2\,\sqrt{j(j + 1)}}{x\,{h}(x) - 2\,\alpha
\,\varepsilon }}\,{y_{4}}(x)=0 
\spazio{1.2}\cr 
&{}&  
{\frac
{d}{dx}}\,{y_{4}}(x) + {\displaystyle \frac {1}{2}} \,(
{\displaystyle \frac {4\,j\,(j + 1)}{x^{2}\,{h}(x)}}  +
{\displaystyle \frac {4\,\alpha ^{2}\,\varepsilon ^{2} - x^{2}\,
{h}^2(x)}{x\,(x\,{h}(x) - 2\,\alpha \,\varepsilon
 )}} )\,{y_{2}}(x) - 
  {\displaystyle \frac
{2\,\sqrt{j(j + 1)}\,{y_{3}}(x)}{x\,{h}(x)}} + \frac 1x y_{4}(x)=0\cr
&{}& 
\label{sysdispari}
\end{eqnarray}
where $h(x)=2+\alpha^2 w/2+\alpha/x$. The system
decouples only for $j=0$ so that, for the moment, we will assume
$j>0$ (remark that the triplet ground state has $j=1$) .

A superficial check of the coefficients shows the presence of an irrelevant
singularity at positive values of $x$ for $\varepsilon>1/2$. The
situation, however, is not so simple and there actually exists a
further hidden singularity  that reveals itself when trying to integrate
numerically the system in a direct way. This unexpected singularity
becomes manifest when deducing the fourth order differential equation equivalent to the system (\ref{sysdispari}) . Both the
partial and the final results when obtaining  this equation are
very cumbersome: the final analytic expression have
been calculated and managed only by means of a systematic use of
computer algebra and cannot be reported here. We
simply give the steps of the method we have followed to get the
equation and that in mathematics is known as the prolongation
method. First we isolate $y_2(x)$ from the first equation and
$y_4(x)$ from the third; we then substitute $y_2(x)$, its
derivative and $y_4(x)$  in the second equation, obtaining a
second order equation in $y_1(x)$  where  $y_3(x)$  and its first
derivative only appear.  We then substitute $y_4(x)$, its
derivative and $y_2(x)$  in the fourth equation, obtaining a
second order equation in $y_3(x)$  where  $y_1(x)$  and its first
derivative only appear. We next consider the prolongations of this
system: this means that we differentiate these two second order
equations and substitute the second order derivatives obtained by
the second order equation themselves. The third order equation for
$y_1(x)$ , thus, contains $y_3(x)$ and its first derivative only.
We differentiate once more this equation and finally from this,
the two third order and the two second order equations we can
eliminate $y_3(x)$ and its first, second and third derivatives,
finally obtaining a fourth order differential equation for
$y_1(x)$.

In the  denominator of the coefficients of the fourth order
equation we find a factor responsible for the appearance of the
new singularity given by a root of the
corresponding equation: 
\begin{eqnarray}
&{}&\!\!\!\!\!\!\!\!\!\! - (\alpha ^{2}\,w + 4)\,(16 + \alpha ^{2}\,w\,(\varepsilon
^{4} \,\alpha ^{2} + 1)\,(\alpha ^{2}\,w + 8))\,x^{3} - 2\,\alpha
(\varepsilon ^{4}\,\alpha ^{4}\,w\,(\alpha
^{2}\,w + 8)\,(3 + 2\, \varepsilon )
\spazio{0.6}\cr 
&{}&\!\!\!\!\!\!\!\!\!\!  
 - (\alpha ^{2}\,w +
4)^{2}\,(2\,\varepsilon  - 3)
+ 32\,\varepsilon ^{4}\,\alpha ^{2})\,x^{2}  + 4\,\alpha
^{2}\,( - \varepsilon ^{4}\,(4\,\varepsilon
 + 3)\,\alpha ^{2} - 3 - 4\,\varepsilon \,(\varepsilon ^{2} -
\varepsilon  - 1)   
\spazio{0.6}\cr 
&{}& \!\!\!\!\!\!\!\!\!\! 
+ 4\,\varepsilon
^{4}\,j(j + 1))(\alpha ^{2}\,w + 4)\,x + 8\,\alpha ^{3}\,( 
4\,\varepsilon ^{4}\,j(j + 1) - (2\,
\varepsilon  + 1)\,( 2\,\varepsilon-1 )^{2}-
\varepsilon ^{4}\,(2\,\varepsilon
 + 1)\,\alpha ^{2})=0\,.\cr
&{}& 
\end{eqnarray}
For the typical values $j=1$ and $w=-0.5$ it has a
solution $x>0$ for $\varepsilon\simeq 0.36014$. Here also we
stress that the presence of this singularity does not affect the
perturbative approach to the hyperfine interaction, but must be
considered if one is willing to integrate the odd equation as it stands.

We have solved the fourth order equation for small enough values
of $~\varepsilon$ --  so to prevent the presence of the additional
singularity -- as well as for $~\varepsilon=1/2$. In both cases,
taking into account the due differences, the method is
conceptually a variant of the double shooting procedure described
in the previous paragraph. For small values of  $\varepsilon$ we
have to take care of the origin and infinity, that are the only
two singular points. It turns out that in each one of them there
exist two regular solutions that can provide the necessary initial
conditions for starting the numerical integration. The spectral
condition is now given by the matching of function, first, second
and third derivatives in a fixed crossing point $x_c$, so to
recover the ``classical solution'' of the equation. This amounts to
the vanishing of the following determinant: 
\begin{eqnarray}
{\det} \left( \! {\begin{array}{cccc}
{y_{0,1}}({x_c}) & y_{0,2}({x_c}) & y_{\infty,1}({x_c}) &
y_{\infty,2}({x_c})
\spazio {1.4}  \\
y_{0,1}^{\,\mathrm{(I)}}({x_c}) & y_{0,2}^{\,\mathrm{(I)}}({x_c})
& y_{\infty,1}^{\,\mathrm{(I)}}({x_c}) &
y_{\infty,2}^{\,\mathrm{(I)}}({x_c})
\spazio {1.4}  \\
y_{0,1}^{\,\mathrm{(II)}}({x_c}) &
y_{0,2}^{\,\mathrm{(II)}}({x_c}) &
y_{\infty,1}^{\,\mathrm{(II)}}({x_c}) &
y_{\infty,2}^{\,\mathrm{(II)}}({x_c})
\spazio {1.4}  \\
y_{0,1}^{\,\mathrm{(III)}}({x_c}) &
y_{0,2}^{\,\mathrm{(III)}}({x_c}) &
y_{\infty,1}^{\,\mathrm{(III)}}({x_c}) &
y_{\infty,2}^{\,\mathrm{(III)}}({x_c})
\end{array}}
  \! \right)\! =0\cr
\label{determinante}
\end{eqnarray}
where $y_{0,i},y_{\infty,i}$, $(i=1,2)$, are the two
regular solutions coming from the origin and from the infinity
respectively and the superscripts denote the order of the
derivatives. As in the even case, the two acceptable series solutions
in the neighborhood of the origin are of the form  $y_{i}(x)=A_i
\,x^{\rho_i}S_i(x)$, where, for $j>0$ and $\varepsilon<1/2$,
\begin{eqnarray}
&{}& \rho_1=  - 1 + \frac 1{2\,(1 -
2\,\varepsilon )} \Bigl[ (1 - 2\,\varepsilon )^{2}\,(4
 - \alpha ^{2}\,(4\,\varepsilon  + 1))  -
4\,( - 1 + 2\, \varepsilon )\,j(j + 1)\Bigr]^{1/2}\cr
&{}& \rho_2=\frac 12 \Bigl[ - \alpha
^{2}\,(2\,\varepsilon  + 1) ^{2} + 4\,(2\,\varepsilon  + 1)\,j(j +
1)\Bigr]^{1/2}
\end{eqnarray}
The search for the asymptotic solutions is a bit more
delicate. Indeed they are still of the form $y_i(x)=B_i\,
\exp[-\kappa_i x]\,x^{\nu_i}T_i(x)$, $i=1,2$, and again
$\kappa_1=\kappa_2=\kappa$, with 
\begin{eqnarray}
\kappa=\frac \alpha 4 \,\Bigl[ - w\,(8+\alpha ^{2}\,w)\Bigr]^{1/2}
\label{cappaeq4}
\end{eqnarray}
while the indices are 
\begin{eqnarray}
&{}& \!\!\!\!\!\!\! \!\!\!\!\!\!\! \!\!\!\!\!\!\! \nu_1=-1 +
{\displaystyle \frac {16+\alpha ^{2}\,w\,(8+\alpha ^{2}\,w
)\,(2\,\varepsilon  + 1)}{2\,(4+\alpha
^{2}\,w)\,\Bigl[ - w\,(8+\alpha
^{2}\,w)\Bigr]^{1/2}}}\spazio{0.8}\cr &{}& \!\!\!\!\!\!\!
\!\!\!\!\!\!\! \!\!\!\!\!\!\! \nu_2=\nu_1-2
\label{nueq4}
\end{eqnarray}
Although an integer difference of the indices could imply
solutions of different type, the present case is the simplest one
and a second non logarithmic solution is found.

We now consider the odd ground state for $\varepsilon=1/2$. We
have then to discuss the fourth order differential equation with
$j=1$ in the presence of a singularity located, almost
independently of the value of $w$, around $x\simeq0.0016478$. The
two regular asymptotic solutions are of the type already
described and the corresponding parameters are obtained from 
(\ref{cappaeq4}) and (\ref{nueq4}) with $\varepsilon=1/2$.
The second solution is again non
logarithmic. The behavior in the origin, however is here
different. In fact we have one of the two regular solutions of the
form $y_1(x)=x^{\sqrt{4-\alpha^2}}S_1(x)$, but the second one, due
to the fact that the singularity in the origin is irregular, must
be searched in  a more general form \cite{Cope}, and results in
\begin{eqnarray}
y_2(x)=\exp\Bigl[  -\frac{4\,\sqrt{\alpha(4+\alpha^2
w)}}{(4+\alpha^2 w)\,\sqrt x} \Bigr]\,x^{-3/4}\,S_2(\sqrt x)
\end{eqnarray}
where $S_2(\sqrt x)$ is a power series in $\sqrt x$ with
zeroth order term equal to unity. Finally four regular solutions
are found in a neighborhood of the singular point: their indices
are $0,1,2,49/\!16$, but despite the integer differences all of them
are non logarithmic. These solutions together with their first
three derivatives are used to bridge the two solutions from the
origin to the two solutions from infinity. From a numerical point,
however, this is not very simple due to the critical sensitivity of
the coefficients of the differential equation to tiny variations
of the coordinate $x$. Indeed the ordinary integration codes
present errors too large to be accepted, so that, in order not to
lose in accuracy when integrating out of the singularity, we have
chosen a mash of sufficiently close points (all of them obviously
regular with respect to the differential equations):  for each
point we have constructed four regular series solutions with a
number of terms sufficiently large to respect the accuracy
requirements, and we have connected all these solutions by
matching functions and derivatives up to order three.

We finally discuss the spectral solution of the odd case
with $j=0$. We have already stated that now the system decouples
and gives rise to a pair of second order differential equations.
As in the even case, only one of these equations, namely the
equation coming from the $(y_1,y_3)$ subsystems, has a physical
meaning. From (\ref{sysdispari}) we easily find
\begin{eqnarray}
&{}&\!\!\!\!\!\!\!\!\!\! {\frac {d^{2}}{dx^{2}}}\,{y}(x) + \frac 2x \Bigl[1 +
{\displaystyle \frac {\alpha }{\alpha ^{2}\,w\,x + 2\, \alpha  +
8\,x}}  - {\displaystyle \frac {\alpha }{4\,x + \alpha ^{2}\,w\,x
+ 2\,\alpha }}+ {\displaystyle \frac {1}{x\,\alpha
\,w + 2}} \Bigr] \,{\frac {d}{dx}}\,{y}(x)
 \spazio{1.4}\cr
 &{}&\!\!\!\!\!\!\!\!\!\!  + 
\Bigl[{\displaystyle \frac {\alpha ^{2}\,w\,(\alpha ^{2}\,w + 8)}{16}} 
+ {\displaystyle \frac {\alpha ^{2}\,(
4\,\varepsilon  + 1)}{4\,x^{2}}}+ {\displaystyle \frac {\alpha \,(\alpha ^{2}\,w + 4)\,(2\,
\varepsilon  + 1)}{4\,x}}    
- {\displaystyle \frac {8\,\alpha \,\varepsilon }{(4 + 
\alpha ^{2}\,w)\,x + 2\,\alpha }}\Bigr]\,y(x)=0\cr
&{}&
\end{eqnarray}
and we see that the coefficient of the first derivative
has a singularity at the point $x=-2/(w\alpha)$, independent of
$\varepsilon$. Since for the bound states we are studying $w$
assumes negative values, the singularity is therefore located at a
finite positive value of $x$ and must be accounted for in the integration
for any value of $\varepsilon$. The situation is similar to
what we have already seen and may briefly summarized as follows:
the index for the acceptable series solution in the origin is
$\rho=-1+\frac 12\,\sqrt{4-\alpha^2(4\varepsilon+1)}$. The two
constants of the asymptotic solution are the same $\kappa$ as in (\ref{cappaeq4})
and $\nu=\nu_1$ as given in (\ref{nueq4}).
In the neighborhood of the singular point the indices are
$0,2$ and there exist two non logarithmic solutions.


\sect{Discussion of the results} \label{Results}

To begin the discussion of the results we report the values we have 
obtained in \cite{RGES} for the levels of the pure Coulomb interacting system.
The classification scheme we used in that paper was fit to describe the spectral terms
for systems with variable mass ratio. 
For convenience in comparing our levels with the corresponding
values existing in literature \cite{Berko,FM} and obtained by semi-classical expansions, we will
adopt the commonly accepted classification of the Positronium levels, 
\footnote {Although not necessary for the discussion,
it is straighforward to recognize the following correspondence between \cite{RGES} and \cite{Berko}:
$(+,0,I)\rightarrow 1^1s_0$, $(-,1,I)\rightarrow 1^3s_1$, $(+,0,II)\rightarrow 2^1s_0$, $(+,1,I)\rightarrow 2^3p_1$, $(+,1,II)\rightarrow 2^1p_1$, $(-,0,I)\rightarrow 2^3p_0$, $(-,1,II)\rightarrow 2^3s_1$, $(-,2,I)\rightarrow 2^3p_2$.
}. 

Our results for the relativistic two body equation with pure Coulomb interaction are as follows. For the
ground states we have
\bigskip
\begin{center}
\begin{tabular}{|c|c|c|c|}
  \hline
  $\phantom{x}$State$\phantom{{}^{{}^{\displaystyle{i}}}}$
  &$w_{\,\mathrm{Coulomb}}$ & $\phantom{x}$State$\phantom{{}^{{}^{\displaystyle{i}}}}$
  & $w_{\,\mathrm{Coulomb}}$
  \\
  \hline
  $1^1s_0\phantom{{}^{{}^{\displaystyle{i}}}}$
  &$\!\,$ -.4999950109 & $1^3s_1\phantom{{}^{{}^{\displaystyle{i}}}}$
  &$\!\,$ -.4999950109 $\spazio{0.5}$\\ \hline
  \end{tabular}
\end{center}
\bigskip

\noindent
For first excited states, the data are:
\bigskip
\begin{center}
\begin{tabular}{|c|c|c|c|}
  \hline
  $\phantom{x}$State$\phantom{{}^{{}^{\displaystyle{i}}}}$
  &$w_{\,\mathrm{Coulomb}}$ & $\phantom{x}$State$\phantom{{}^{{}^{\displaystyle{i}}}}$
  & $w_{\,\mathrm{Coulomb}}$
  \\
  \hline
  $2^1s_0\phantom{{}^{{}^{\displaystyle{i}}}}$
  &$\!\,$ -.1249996884
  & $2^3s_1\phantom{{}^{{}^{\displaystyle{i}}}}$
  &$\!\,$ -.1249996884 $\spazio{0.5}$
  \\
  \hline
  $2^1p_1\phantom{{}^{{}^{\displaystyle{i}}}}$
  &$\!\,$ -.1250002427
  & $2^3p_1\phantom{{}^{{}^{\displaystyle{i}}}}$
  &$\!\,$ -.1250005200 $\spazio{0.5}$
  \\
  \hline
  $2^3p_0\phantom{{}^{{}^{\displaystyle{i}}}}$
  &$\!\,$ -.1250007974
  & $2^3p_2\phantom{{}^{{}^{\displaystyle{i}}}}$
  &$\!\,$ -.1249999654 $\spazio{0.5}$
  \\ \hline
  \end{tabular}
\end{center}

\bigskip

As explained in Appendix A, the levels accounting for the first order
perturbative effects of the Breit terms are given by 

\begin{eqnarray}
w_{\,\mathrm{Breit}}=w_{\,\mathrm{Coulomb}}+\frac
12\,\frac{dw(\varepsilon)}{d\varepsilon}\Bigr|_{\,\varepsilon=0}
\end{eqnarray}

The derivative of $w$ in $\varepsilon=0$ was calculated  by computing  the eigenvalues corresponding to $\varepsilon=0.1$ and $\varepsilon=0.2$ and 
then looking for the three point Lagrangian interpolation through them and through  $(\varepsilon=0,w=w_{\mathrm{Coulomb}})$. The results has been checked by repeating the same procedure for other values of $\varepsilon$ sufficiently close to the origin and the differences are at most 
of some unities on the last figure for the states $s$ and even less for the states $p$.  Even if we take the values
we have calculated for $\varepsilon=1/2$ -- and reported below -- to construct the interpolation, we see that the results for the hyperfine ground levels differ only for some unities on the ninth figure. 
In the following table we summarize the data we have obtained. Remark that the calculations have been done with a number of figures sufficiently large to prevent the rounding errors.
\bigskip
 \begin{center}
\begin{tabular}{|c|c|c|c|}
  \hline
  $\phantom{x}$State$\phantom{{}^{{}^{\displaystyle{i}}}}$
  &$w_{\,\varepsilon=0.1}$ & $w_{\,\varepsilon=0.2}$ & $dw/d\varepsilon|_{\,\varepsilon=0}$ $\spazio{0.5}$
  \\
  \hline
  $1^1s_0\phantom{{}^{{}^{\displaystyle{i}}}}$
  &$\!\,$ -.5000008658 &$\!\,$ -.5000024624
  &$\!\,$ -.7984077369$\cdot 10^{-4}$ $\spazio{0.5}$\\ \hline
  $1^3s_1\phantom{{}^{{}^{\displaystyle{i}}}}$
  & $\!\,$-.4999955437 &$\!\,$ -.4999953671
  &$\!\,$ -.8874896904$\cdot 10^{-5}$  $\spazio{0.5}$ 
\\ \hline
  $2^1s_0\phantom{{}^{{}^{\displaystyle{i}}}}$
  & $\!\,$-.1250005867 &$\!\,$ -.1250009527
  &$\!\,$ -.1164413908$\cdot 10^{-4}$ $\spazio{0.5}$
  \\
  \hline
  $2^3s_1\phantom{{}^{{}^{\displaystyle{i}}}}$
  &$\!\,$ -.1249999215 &$\!\,$ -.1250000659
  & $\!\,$-.2774401032$\cdot 10^{-5}$  $\spazio{0.5}$
  \\
  \hline
  $2^1p_1\phantom{{}^{{}^{\displaystyle{i}}}}$
  &$\!\,$ -.1250003204 & $\!\,$-.1250002205
  & $\!\,$-.1664091904$\cdot 10^{-5}$  $\spazio{0.5}$
  \\
  \hline
  $2^3p_1\phantom{{}^{{}^{\displaystyle{i}}}}$
  &$\!\,$ -.1250007197 &$\!\,$ -.1250008750
  &$\!\,$ -.2218611453$\cdot 10^{-5}$  $\spazio{0.5}$
  \\
  \hline
  $2^3p_0\phantom{{}^{{}^{\displaystyle{i}}}}$
  &$\!\,$ -.1250012966 & $\!\,$-.1250017959
  &$\!\,$ -.4992271268$\cdot 10^{-5}$  $\spazio{0.5}$
  \\
  \hline
  $2^3p_2\phantom{{}^{{}^{\displaystyle{i}}}}$
  &$\!\,$ -.1250000186 &$\!\,$ -.1250000452
  & $\!\,$-.6646165115$\cdot 10^{-6}$  $\spazio{0.5}$
  \\
  \hline
\end{tabular}
\end{center}

\bigskip

The spectral values with  $\varepsilon=1/2$ for $1^1s_0$ and $1^3s_1$ are the following:
\bigskip
\begin{center}
\begin{tabular}{|c|c|c|c|}
  \hline
  $\phantom{x}$State$\phantom{{}^{{}^{\displaystyle{i}}}}$
  &$w_{\,\varepsilon=1/2}$ & $\phantom{x}$State$\phantom{{}^{{}^{\displaystyle{i}}}}$
  & $w_{\,\varepsilon=1/2}$
  \\
  \hline
  $1^1s_0\phantom{{}^{{}^{\displaystyle{i}}}}$
  &$\!\,$ -.4999816915 & $1^3s_1\phantom{{}^{{}^{\displaystyle{i}}}}$
  &$\!\,$ -.4999905793 $\spazio{0.5}$\\ \hline
  \end{tabular}
\end{center}

\bigskip

The differential equations producing these last values present a singularity at finite values of the radial coordinate.  This singularity is in fact non disturbing for the integration  
and one can ``pass through''  by imposing the matching conditions as explained in Section III. The results however confirms the Breit idea: indeed they present a qualitatively wrong configuration, with the singlet  higher than the triplet.

Let us now compare our results with known data. We remind the formula giving the first terms of the semi-classical expansion for the singlets \cite{Landauetal}:
\begin{eqnarray}
w=-\frac 1{2 n^2}+\frac {\alpha^2}{2 n^4}\Bigl(\frac{11}{16}-\frac n{j+\frac 12}\Bigr) +O(\alpha^4)
\end{eqnarray}
It is well known that the $1^3s_1$ level has a value 
\begin{eqnarray}
w=-\frac 1{2}+\frac {1}{96}\alpha^2
\end{eqnarray}
omitting the annihilation term, that contributes for an additional $\alpha^2/2$ to the singlet-triplet splitting.
From \cite{Berko} we take the approximations for the $n=2$ triplets obtained from \cite{FM}. The results are
\begin{eqnarray}
w(2^3p_0)=-\frac 1{8}+\frac {95}{1536}\alpha^2
\end{eqnarray}
and 
\begin{eqnarray}
\!\!\!\!\!  w(2^3p_j)=w(2^3p_0)+\delta(j)\,, ~~~w(2^1p_1)=w(2^3p_0)+\delta'
\end{eqnarray}
with
\begin{eqnarray}
\delta(1)=\frac 5{160}\alpha^2,~~~\delta(2)=\frac 9{160}\alpha^2,~~~\delta'=\frac 1{24}\alpha^2.
\end{eqnarray}
Finally $w(2^3s_1)$ is deduced from the relation  \cite{KK}
\begin{eqnarray}
w(2^3s_1)-w(2^1s_0)=\frac 18\, \Bigl(w(1^3s_1)-w(1^1s_0)\Bigr)
\end{eqnarray}
The comparison with our results is summarized in the following table:
\bigskip
 \begin{center}
\begin{tabular}{|c|c|c|}
  \hline
  $\phantom{x}$State$\phantom{{}^{{}^{\displaystyle{i}}}}$ &$w_{\,\mathrm{num}}$ 
&$w_{\,\mathrm{semi-classical}}$
   \\
  \hline
  $1^1s_0\phantom{{}^{{}^{\displaystyle{i}}}}$
  &$\!\,$ -.5000349313  &
$-{ \frac 12}-{ \frac{21}{32}}\alpha^2\phantom{1}$=-.5000349462$\spazio{0.5}$
 \\ 
\hline
  $1^3s_1\phantom{{}^{{}^{\displaystyle{i}}}}$
  &$\!\,$ -.4999994484  &
$-{ \frac 12}+{ \frac 1{96}}{\alpha^2}\phantom{1}$=-.4999994453$\spazio{0.5}$
 \\ 
\hline
  $2^1s_0\phantom{{}^{{}^{\displaystyle{i}}}}$
  &$\!\,$ -.1250055105 &
$-{ \frac 18}-{ \frac{53}{512}}\alpha^2\,$=-.1250055123$\spazio{0.5}$
 \\ 
  \hline
  $2^3s_1\phantom{{}^{{}^{\displaystyle{i}}}}$
  &$\!\,$ -.1250010756 &
$-{ \frac 18}-{ \frac{31}{1536}}\alpha^2$=-.1250010747$\spazio{0.5}$
 \\ 
\hline
  $2^1p_1\phantom{{}^{{}^{\displaystyle{i}}}}$
  &$\!\,$ -.1250010747  &
$-{ \frac 18}-{ \frac{31}{1536}}\alpha^2$=-.1250010747$\spazio{0.5}$
 \\ 
  \hline
  $2^3p_1\phantom{{}^{{}^{\displaystyle{i}}}}$
  &$\!\,$ -.1250016293  &
$-{ \frac 18}-{ \frac{47}{1536}}\alpha^2$=-.1250016294$\spazio{0.5}$
 \\ 
  \hline
  $2^3p_0\phantom{{}^{{}^{\displaystyle{i}}}}$
  &$\!\,$  -.1250032935 &
$-{ \frac 18}-{ \frac{95}{1536}}\alpha^2$= -.1250032935$\spazio{0.5}$
 \\ 
  \hline
  $2^3p_2\phantom{{}^{{}^{\displaystyle{i}}}}$
  &$\!\,$ -.1250002977  &
$-{ \frac 18}-{ \frac{43}{7680}}\alpha^2$=-.1250002982$\spazio{0.5}$
 \\ 
\hline
  \end{tabular}
\end{center}

\bigskip

A couple of  final comments on the results are in order. In the first place we observe that the 
pure Coulomb levels calculated in the semi-classical approximation or, equivalently, using the expansion in $\alpha$ differ from the levels calculated by the two body relativistic equation by a quantity of the same order of the approximation itself. In order to produce a concrete example we let $\varepsilon=0$ in (\ref{even1}) and we use the ``atomic variable'' $z=\frac 12 \alpha\, x$.  Expanding in $\alpha$ up to the second order, we find
\begin{eqnarray}
\!\!\!\!\!{\frac {d^{2}}{dz^{2}}}\,{y}(z) +  \Bigl(
{\displaystyle \frac {w^{2}\,\alpha ^{2}}{4}}  + 2\,w + 
{\displaystyle \frac {4 +  {\alpha ^{2}\,w}
 }{2z}}  + {\displaystyle \frac {\alpha ^{2}}{4\,z^{2}}} 
 \Bigr) \,{y}(z)
\label{even1_alpha}
\end{eqnarray}
with regular solution
\begin{eqnarray}
&{}& y(z)=z^{\displaystyle{1/2 +(1/2)\, \sqrt{1 - \alpha ^{2}}}}\,e^{ \displaystyle{- (1/2)\, 
\sqrt{ - w\,(\alpha ^{2}\,w + 8)}\,z}}\,\cdot\spazio{0.5}\cr
&{}& \phantom{y(z)=}{}_1F_1\Bigl( 
{\displaystyle \frac {1}{2}}  + {\displaystyle \frac {\sqrt{1 - 
\alpha ^{2}}}{2}}  - {\displaystyle \frac {\alpha ^{2}\,w + 4}{2
\,\sqrt{ - w\,(\alpha ^{2}\,w + 8)}}} , 1 + \sqrt{1 - \alpha 
^{2}},~ 
\sqrt{ - w\,(\alpha ^{2}\,w + 8)}\,z\Bigr)\,.
\label{sol_even1_alpha}
\end{eqnarray}
From the vanishing of the first argument of the hypergeometric we get the ground level
\begin{eqnarray}
&{}&\!\!\!\!\!\!\!\!\!\!w=\frac 1{\alpha^3}\,{\displaystyle \Bigl(2 + 2\,\sqrt{1 - \alpha ^{2}}\Bigr)\,\sqrt{2
 - 2\,\sqrt{1 - \alpha ^{2}}}}  - {\displaystyle 
\frac {4}{\alpha ^{2}}} \simeq -\frac 12-\frac 5{32}\,\alpha^2+O(\alpha^4)
\label{spec_even1_alpha}
\end{eqnarray}
Comparing (\ref{spec_even1_alpha})
with our relativistic pure Coulomb result $w=-.4999950109$, that can be approximated as  $w=-1/2+3\,\alpha^2\!/32+O(\alpha^4)$,  we find 
a difference of about $\alpha^2\!/4$. On the contrary, when looking at the data presented in the final table,
we see that the difference reduces to less than $\alpha^3\!/26$ and becomes even more negligible
for the levels with higher orbital angular momentum.

A second observation concerns the degeneracy of the states $2^3s_1$ and $2^1p_1$ that is predicted from the
perturbative expansion and that is completely confirmed from the numerical calculations. The existing difference of about $0.01\,\alpha^2$ for the pure Coulomb interaction disappears when introduncing the Breit term.
A similar phenomenon in the perturbative framework had been noticed in \cite{Chi} with respect to the Lamb shift in the hydrogen atom, produced by a pure relativistic Coulomb interaction and reabsorbed by the presence of the magnetic term.

To conclude, in this paper we have presented a relativistic calculation of the
hyperfine structure of the Positronium, providing the theoretical and mathematical instruments to obtain the results. We have pursued our approach without any semi-classical approximation
or expansion in the fine structure constant: the only due perturbative treatment has been reserved to the magnetic interaction term.
Along this way, almost unexplored, we have proved some other aside facts. In particular we have discussed the nature and the properties of the singularities arising in the development and we have found that they bear no
serious consequences neither in the integration of the wave equations, nor in their spectral behavior, but for lengthy technical complications: this, in a sense, can be considered an indirect test of the reliability
of the approach to bound states through relativistic wave equations  up to the quantum field theoretic corrections. We have also indicated possible applications of the method, that are now under investigation.


\appendix\section{A perturbative expansion}
\label{pertapp} In this Appendix we prove the relationships
between the derivative of the spectral values with respect to a
parameter $\varepsilon$ and the perturbative expansion in that
parameter.

Consider the Hermitian operators $H$, $Q$ and the sum 
\begin{eqnarray}
K(\varepsilon)=H+\varepsilon Q
\end{eqnarray}
Let $U(\varepsilon)$ be the unitary operator such that
\begin{eqnarray}
K_d(\varepsilon)=U^{-1}(\varepsilon)K(\varepsilon)U(\varepsilon)
\label{Ad}
\end{eqnarray}
is diagonal. Denoting by a dot the derivatives with respect to
$\varepsilon$ and letting $U=U(0)$, $U^{-1}=U^{-1}(0)$, $\dot
U=\dot U(\varepsilon)|_{\varepsilon=0}$, we expand equation
(\ref{Ad}) to the first order in $\varepsilon$ obtaining 
\begin{eqnarray}
K_d+\varepsilon \dot{K}_d=H_d+\varepsilon \Bigl( [H_d,U^{-1}\dot
U]+U^{-1}Q U\Bigr) \label{Ad1}
\end{eqnarray}
where $K_d=K_d(0)$,
$\dot{K}_d=\dot{K}_d(\varepsilon)|_{\varepsilon=0}$. Obviously
$K_d\equiv H_d\equiv U^{-1}HU$ and $\dot{K}_d$ are diagonal
matrices. Moreover the diagonal elements of $[H_d,U^{-1}\dot U]$
are vanishing, so that 
\begin{eqnarray}
(\dot{K}_d)_{ii}=(U^{-1}Q U)_{ii}\equiv\langle
V_i,QV_i\rangle\,, ~~~~~~~ [H_d,U^{-1}\dot
U]_{ij}+(U^{-1}Q U)_{ij}=0\,,~~~~~i\neq j\,, \label{Ad2}
\end{eqnarray}
where $V_i$ is the $i$-th normalized eigenvector of $H$.
Hence 
\begin{eqnarray}
K_d(\varepsilon)_{ii} =(H_d)_{ii}+\varepsilon \langle
V_i,QV_i\rangle +O(\varepsilon^2) \label{FirstOrder}
\end{eqnarray}
and we recover the usual first order correction of the
perturbative expansion.

The procedure goes over to any order. Although in our framework we
do not use anything but the first order, we want just to outline
how the second order is also obtained. A 
straightforward computation gives 
\begin{eqnarray}
\ddot{K}_d=[H_d,U^{-1}\ddot U]+2\,U^{-1}\dot U[U^{-1}\dot
U,H_d]+2\,[U^{-1}Q U,U^{-1}\dot
U] \label{Ad3}
\end{eqnarray}
where, in the usual notations,  $
\ddot{K}_d=d^2K_d(\varepsilon)/d\varepsilon^2|_{\varepsilon=0}$
and the same for $\ddot U$. Again the diagonal part of
$[H_d,U^{-1}\ddot U]$ vanishes. Using the second equation of
(\ref{Ad2}) and expanding the commutators, we find 
\begin{eqnarray}
\!\!\!\!\!\!\! \frac 12
(\ddot{K}_d)_{ii}=\sum_{j\neq i}(U^{-1}\dot U)_{ij}(U^{-1}Q
U)_{ji}+\sum_j(U^{-1}Q
U)_{ij}(U^{-1}\dot U)_{ji} -\sum_{j}(U^{-1}\dot U)_{ij}(U^{-1}Q
U)_{ji} \label{Ad4}
\end{eqnarray}
$\!\!\!$or equivalently, taking carefully into account the
summation ranges, 
\begin{eqnarray}
\frac 12 (\ddot{K}_d)_{ii}=\sum_{j\neq i}(U^{-1}Q
U)_{ij}(U^{-1}\dot U)_{ji}\,.
\end{eqnarray}
Observing that from the second equation of (\ref{Ad2}) we
have 
\begin{eqnarray}
(U^{-1}\dot U)_{ji}=\frac{(U^{-1}Q
U)_{ij}}{(H_d)_{ii}-(H_d)_{jj}}\,,~~~~~j\neq i
\end{eqnarray}
we find the usual second order contribution 
\begin{eqnarray}
\frac {\varepsilon^2}2\,
(\ddot{K}_d)_{ii}=\varepsilon^2\sum_{j\neq i}\frac{|\langle
V_i,QV_j  \rangle|^2}{(H_d)_{ii}-(H_d)_{jj}}\,.
\end{eqnarray}
It appears therefore the advantage achieved for a numerical evaluation of the perturbative corrections
in our problem as previously explained.

\bigskip\bigskip

%
%

\vfill\break


\begin{thebibliography}{999}

\bigskip

\bibitem{Darwin} G.C. Darwin, {Proc. Royal Society} {\bf 118}, 654, (1928).

\bibitem{Breit} Breit, {Phys. Rev.} {\bf 553}, page, (1929).

\bibitem{Fermi} E. Fermi, {Zeitschr. f. Physik} {\bf 60}, 320, (1930).

\bibitem{Pirenne} J. Pirenne,  {Arch. Sci. Phys. Nat.} {\bf 29}, 207, (1947).

\bibitem{Landauetal} V.B. Berestetski, L.D. Landau, {JEPT} {\bf 19}, 673, (1949).

\bibitem{Schwi} J. Schwinger, {Proc. Nat. Acad. Sci. U.S.} {\bf 37}, 452 and 455, (1951).

\bibitem{BetheSalp} H.A. Bethe, E.E. Salpeter , {Phys. Rev.} {\bf 84}, 350, (1951).

\bibitem{Salp}  E.E. Salpeter , {Phys. Rev.} {\bf 87}, 328, (1952).

\bibitem{KK}R. Karplus, A. Klein, {Phys. Rev.} {\bf 87}, 848, (1952).

\bibitem{FM} T. Fulton, P.C. Martin, {Phys. Rev.} {\bf 93}, 904, (1955).

\bibitem{Berko}  S. Berko, H.N. Pendleton; {Ann. Rev. Nuc. Part. Sci.}  {\bf 30}, 543, (1980).

\bibitem{Lepage}  G.P. Lepage, {``\emph{Two body bound states in Quantum Electrodynamics}''}
{Ph.D. Thesis, SLAC 1978}, Microfiche at Fermilab.

\bibitem{Krollinowski}  W. Krollinowski, {Acta Phys. Polonica} {\bf B12}, 981, (1981).

\bibitem{Chi} R.W. Childers, {Phys. Rev.} {\bf D 26}, 2902, (1982).

\bibitem{HG} J.Hata, I.P. Grant, {J. Phys B: At. Mol. Phys.}
{\bf 17}, L107, (1984).

\bibitem{Indelicado}  O. Gorceix, P. Indelicado, J.P. Desclaux, {J. Phys B: At. Mol. Phys.}
{\bf 20}, 639, (1987).

\bibitem{SSM} T.C. Scott, J. Shertzer and R.A. Moore,
{Phys. Rev} {\bf A 45}, 4393, (1992).

\bibitem{Dare} J.W. Darewych and L. Di Leo, {J. Phys. A}, {\bf 29},
6817, (1996); A. Terekidi, J.W. Darewych, {J.Math.Phys}
{\bf 45}, 1474, (2004).

\bibitem{Sazd} J. Mourad, H. Sazdjian, {J. Phys. G}, {\bf 21},
267, (1995).

\bibitem{CL} H. Crater and B. Liu, {Phys. Rev.} {\bf C 67}, 024001,
(2003).

\bibitem{CVA} H. Crater and P. Van Alstine, {Phys. Rev.} {\bf D 70},
034026, (2004).

\bibitem{SZ} E.V. Shuryak, I. Zahed,  hep-ph/0403127, (2004).

\bibitem{Tsibidis}  G.D. Tsibidis, {Acta Physica Polonica} {\bf B35}, 2329, (2004).

\bibitem{Karsh}  S.G. Karshenboim, {Phys. Reports} {\bf 422}, 1, (2005).

\bibitem{Peshkin}  M.E. Peshkin, {``\emph{Aspects of the dynamics of heavy-quark systems}''}
{SLAC Summer Institute 1983}, 151.

\bibitem{Hardekopf} G. Hardekopf, J. Sucher, {Phys. Rev.} {\bf D 33}, 2035, (1986).

\bibitem{RGES} R.Giachetti, E.Sorace, {J. Phys A: Math. and Gen.}
{\bf 38}, 1345, (2005).

\bibitem{OrazioWa} H.W. Crater, C.W. Wong, C.Y. Wong {Int. J. Mod. Phys.} {\bf E 5}, 589, (1996).

\bibitem{Orazio} H.W. Crater,  {J. Comput. Phys.} {\bf D 115}, 470, (1994).

\bibitem{AhB} A.I. Akhiezer, V.B. Berestetski, ``\textit{Quantum Electrodynamics}'', Wiley Interscience Publishers, (New York, 1965).



\bibitem{PartData} ``Review of Particle Physics'', {Phys. Lett.}, {\bf B592}, 1, (2004)

\bibitem{CoddLev} E.A. Coddington, N.  Levinson, ``\textit{Theory of Ordinary Differential Equations}'', McGraw-Hill Book Co., (New York, 1955).

\bibitem{Cheon} I. Tsutsui, T. F\"ul\"op, T. Cheon,
arXiv preprint quant-ph/0209110 (2002).

\bibitem{Cope} F.T. Cope, {Amer. J. Math.} {\bf 56}, 411 (1934) and
{\bf 58}, 130 (1936).

\end{thebibliography}
\end{document}